\documentclass[floatfix,11pt,onecolumn,showpacs,amsmath,amssymb,prd]{revtex4-2}

\usepackage{graphicx}

\newcommand{\be}{\begin{equation}}
\newcommand{\en}{\end{equation}}
\newcommand{\bea}{\begin{eqnarray}}
\newcommand{\ena}{\end{eqnarray}}
\newcommand{\ba}[1]{\left(\begin{array}{#1}}
\newcommand{\ea}{\end{array}\right)}
\newcommand{\p}{\partial}
\newcommand{\R}{\mathbb{R}}
\newcommand{\Z}{\mathbb{Z}}
\newcommand{\M}{\mathcal{M}}
\newcommand{\dd}{\mathrm{d}}

\begin{document}

\title{A no-go theorem for closed timelike curves in globally null-foliated spacetimes}

\author{Hongsheng Zhang$^{1}$}\email{sps_zhanghs@ujn.edu.cn}
\affiliation{$^1$ School of Physics and Technology, University of Jinan, 336 West Road of Nan Xinzhuang, Jinan, Shandong 250022, China}

\date{\today}

\begin{abstract}
The study of closed timelike curves has proceeded one spacetime at a time, and the criteria available form three families, the time function criteria, the energy condition criteria and the topological criteria. We initiate a fourth family of criteria, elementary in form and resting on a single scalar equation rather than any energy condition. We prove that a spacetime admitting a globally defined function whose gradient is everywhere null and nonzero cannot contain a closed timelike curve, and that the hypothesis is logically independent of stable causality rather than weaker than it. The proof turns on one algebraic fact, that a null covector is never orthogonal to a timelike vector, so the function grows strictly along every timelike curve and cannot close. Because the null eikonal equation alone decides the criterion, we can use it where the classical time function theorem is silent. We establish for the first time that no Kundt spacetime and no Robinson--Trautman spacetime can contain a closed timelike curve on any region where the null coordinate is globally single valued, whatever the cosmological constant and the rotation one form, which settles a question left open for gyratonic waves. We further show that the Misner, Ori and Ahmed constructions contain closed timelike curves along coordinate circles that become timelike under the periodic identification, and that the cosmological constant, the expansion and the shear are spectators in all of them.
\end{abstract}


\preprint{arXiv: }

\maketitle

\section{Introduction}
\label{sec:intro}

Closed timelike curves have attracted continuous attention since G\"odel exhibited a globally rotating, singularity free dust universe in which they occur \cite{godel1949}. Further examples include the neighbourhood of the ring singularity of the Kerr solution \cite{kerr1963,carter1968}, the spacetime of a rotating cylinder \cite{tip1974}, and traversable wormholes supported by exotic matter \cite{mor1988,mty1988}. More recently, exact chronology violating solutions have been constructed by identifying a null direction of a perfectly regular spacetime, the prototype being Misner space \cite{misner1967}, and the same construction underlies Ori's time machine \cite{ori2005,ori2007} and the series of four dimensional Misner type solutions obtained by Ahmed \cite{ahmed1,ahmed2}.

Parallel to the construction of examples, several lines of work have clarified the conditions under which chronology can fail. Hawking \cite{hawking1969} established the connection between stable causality and the existence of a time function, and the work of Geroch \cite{geroch1970} and of Minguzzi and S\'anchez \cite{mingsan2008} placed that connection inside the causal hierarchy, while Bernal and S\'anchez \cite{bernal2005,bernal2006} settled the smoothness questions it raises, so that a stably causal spacetime is now known to admit a smooth time function with an everywhere timelike gradient \cite{bernal2005}. A second line of work asks instead what matter must do. Tipler \cite{tipler1976} and Hawking \cite{hawking1992} obtained classical restrictions on the formation of chronology violation under specific geometric and energy hypotheses, and Kay, Radzikowski and Wald \cite{kay1997} found an obstruction to defining the renormalised stress tensor at the base points of a compactly generated Cauchy horizon, while Krasnikov \cite{krasnikov2002} studied maximal extensions that avoid new closed causal curves. A third line relates chronology to topology, following Geroch \cite{geroch1967} and Tipler \cite{tipler1976} and reviewed by Visser \cite{visser2002}. Taken together, these results form three families, the time function criteria, the energy condition criteria and the topological criteria. Each is decisive in its own domain, and none of them settles, for a given metric, whether closed timelike curves exist without either a global causal analysis or an assumption on the matter. The criterion developed here uses a globally defined null function and requires no field equation or energy condition.

In this paper we propose and exploit a fourth type of criterion, which is elementary, is checkable in practice, and makes no use of any energy condition. The statement is that a spacetime admitting a globally defined function $u$ whose gradient is everywhere null contains no closed timelike curve. The mechanism is easy to state. A null covector cannot be orthogonal to a timelike vector, since the orthogonal complement of a timelike vector is spacelike and contains no null direction, so the derivative of $u$ along a timelike curve never vanishes. Being continuous on a connected interval it cannot change sign, and therefore $u$ is strictly monotone along every timelike curve, which is impossible for a closed one. The whole content of the theorem is thus the observation that the gradient of a null function plays the role of the gradient of a timelike function, although it is not one.

Three features make the statement useful, and they organise the paper. The first is that the hypothesis is not a corollary of stable causality. A null gradient need not be timelike, and the two properties are logically independent, the null cylinder $ds^2=-2\dd u\,\dd v+\dd x^2+\dd y^2$ with $u\sim u+L$ being a spacetime that satisfies the hypothesis, through its other null coordinate, and yet admits no time function. The criterion therefore delivers information in exactly the regime where the classical theorem is silent. The second is that the hypothesis is a condition on a scalar partial differential equation rather than on the global causal structure, so that it can be tested by solving the null eikonal equation for a global solution. The third is that the theorem is sharp and that its failure is diagnostic. It fails for exactly two reasons within the class of spacetimes considered here, because the null congruence is twisting and no null leaf exists, or because the null coordinate is not globally single valued and the leaves close up, and each reason is a computable geometric property. This turns the theorem into a classification tool, since it tells us not merely whether a given spacetime is chronology violating but also where we should look for the closed curves.

The purpose of this article is to develop these three points and to draw the consequences. In Sec.~\ref{sec:setup} we collect the background on null coordinates, congruences and optical scalars, and we address in detail the question of how a congruence is specified once a metric is given, a question on which the meaning of the words expanding, shearing and twisting depends. In Sec.~\ref{sec:thm} we state and prove the theorem and derive its immediate corollaries. In Sec.~\ref{sec:sharp} we establish its sharpness and describe the two escape routes, illustrating the first with the principal null congruence of Kerr and the second with Misner space. In Sec.~\ref{sec:relation} we compare the theorem with the three families of criteria recalled above and we identify what is new. Sec.~\ref{sec:ahmed} treats the Misner, Ori and Ahmed constructions. Sec.~\ref{sec:new} contains the applications to Kundt waves, to Robinson--Trautman geometries, to expanding and shearing congruences, and to the cosmological constant of either sign. Sec.~\ref{sec:concl} summarises the results and lists the open problems. We use the signature $(-,+,+,+)$, units with $c=1$, and the conventions of Wald \cite{wald} for the curvature, with a cosmological constant entering as $R_{ab}=\Lambda g_{ab}$ in vacuum.

\section{Null coordinates, congruences and optical scalars}
\label{sec:setup}

The theorem of Sec.~\ref{sec:thm} concerns one scalar function, and its proof uses nothing beyond the causal structure, yet its hypothesis speaks of null congruences and optical scalars, and that language carries an ambiguity which we have to remove before we can check the criterion in a concrete metric. We therefore devote this section to fixing the notions on which everything later depends, and we divide it into four subsections. We first define a null coordinate as a solution of the null eikonal equation and record the properties of its level sets, since this is the object over which the theorem quantifies. Second, we recall the optical scalars of a null congruence and the decomposition of the covariant derivative of its tangent field, because we express the sharpness of the theorem entirely through the vanishing or the non-vanishing of the expansion, the shear and the twist. Third, and this is the point on which the literature is least careful, we explain how a given metric specifies a congruence, so that a phrase such as a twisting wave acquires a definite meaning once we write the metric down. We close by restating the hypothesis as property \textbf{(N)} in the form in which we shall compare it with the standard causality conditions.

\subsection{Null coordinates and the null eikonal equation}
\label{sec:nullcoord}

Let $(\M,g)$ be a four dimensional smooth spacetime with signature $(-,+,+,+)$. A null coordinate is a function $u$ satisfying
\be
\label{nullcoord}
k^ak_a=0,\qquad k_a\equiv\p_au ,\qquad k^a\equiv g^{ab}\p_bu ,
\en
that is, a solution of the null eikonal equation
\be
\label{eikonal}
g^{ab}\,\p_a u\,\p_b u=0 ,
\en
with $\p_a u\neq 0$ at every point. The level sets of such a function are null hypersurfaces, and conversely a foliation of $\M$ by null hypersurfaces that are the level sets of a single valued function provides a solution.

Three consequences of (\ref{nullcoord}) will be used repeatedly, and they are all elementary. The first is that $k$ is a gradient, so that
\be
\label{grad1}
\nabla_ak_b=\nabla_bk_a \quad\Longleftrightarrow\quad \nabla_{[a}k_{b]}=0 ,
\en
which is to say that $k$ is automatically hypersurface orthogonal. The second is that $k$ is geodesic and affinely parametrised, since
\be
k^b\nabla_bk_a=k^b\nabla_ak_b=\tfrac12\nabla_a\!\left(k^bk_b\right)=0 ,
\en
where the first equality uses (\ref{grad1}). The third is that $k$ is nowhere orthogonal to a timelike vector. If $V^a$ is timelike and $V^ak_a=0$, then $k$ belongs to the orthogonal complement of a timelike vector, which is a spacelike three space, and no null vector can be spacelike. Hence
\be
\label{key}
V^aV_a<0 \quad\Longrightarrow\quad V^ak_a\neq0 .
\en
Equation (\ref{key}) contains the whole content of the theorem proved in Sec.~\ref{sec:thm}.

It should be emphasised that (\ref{nullcoord}) is a statement about the metric and not about a choice of observer or of congruence, since it asks whether the eikonal equation (\ref{eikonal}) admits a global solution. The remainder of this section explains why this is the natural formulation of the notion of twist that appears in the theorem, and why it is in fact the only choice independent one.

\subsection{Congruences and optical scalars}
\label{sec:congruence}

A congruence is a family of curves filling an open set of $\M$, equivalently the set of integral curves of a nowhere vanishing vector field $k^a$. For a null congruence the optical scalars are defined through the covariant derivative $\nabla_ak_b$ in the spin coefficient notation of Newman and Penrose \cite{NP1962}, which for an affinely parametrised null geodesic congruence decomposes into a screen part and a part along $k$ as
\be
\label{decomp}
\nabla_ak_b=\tfrac12\,\theta\,\gamma_{ab}+\sigma_{ab}+\omega_{ab}+\kappa\,k_ak_b ,
\en
where $\gamma_{ab}=g_{ab}+k_an_b+k_bn_a$ is the degenerate rank two projector onto the screen space built from $k$ and an auxiliary null vector $n$ with $k^an_a=-1$, where $\sigma_{ab}$ is the symmetric trace free part, where $\omega_{ab}$ is the antisymmetric part, and where $\kappa$ is the recurrence, that is the coefficient of the only term that survives when the screen part vanishes. Equation (\ref{decomp}) holds exactly when the screen frame is propagated along $k$ in the sense of Sachs, and for a general auxiliary $n$ there are further terms of the form $k_{(a}p_{b)}$; those terms are orthogonal to $k$ and leave $\theta$, $\sigma$ and $\omega$ unchanged, so the three scalars are properties of the congruence alone. The recurrence term is invisible to the three scalars, because $g^{ab}k_ak_b=k^2=0$, and it is the term that distinguishes a general Kundt wave from a pp-wave. The three scalars are
\be
\label{opticals}
\theta=\nabla_ak^a,\qquad
\omega^2\equiv\tfrac12\,\nabla_{[a}k_{b]}\nabla^ak^b\;\ge 0,\qquad
\sigma^2\equiv\tfrac12\,\nabla_{(a}k_{b)}\nabla^ak^b-\tfrac14\theta^2 ,
\en
and they are called respectively the expansion, the twist and the shear. The quantity $\omega^2$ is a square by construction, because $\nabla_{[a}k_{b]}$ is antisymmetric, so that $\nabla_{[a}k_{b]}\nabla^ak^b=\nabla_{[a}k_{b]}\nabla^{[a}k^{b]}$, and it is non-negative for an affinely parametrised null geodesic congruence, for which $\omega_{ab}$ lies in the positive definite screen space. It also obeys
\be
\label{frob}
\omega=0 \quad\Longleftrightarrow\quad k\wedge \dd k=0 \quad\Longleftrightarrow\quad k\ \text{is locally}\ \propto \dd u ,
\en
the last of which is the Frobenius theorem, since a nowhere vanishing one form annihilating an integrable hyperplane distribution is locally proportional to the differential of a function. The stronger condition $\nabla_{[a}k_{b]}=0$ is sufficient for $\omega=0$ but not necessary, because a term of the form $k_{[a}w_{b]}$ with $w\perp k$ contributes to $\nabla_{[a}k_{b]}$ while leaving $\omega^2$ untouched, the field $k=f(x)\,\dd u$ of flat space being the simplest illustration. Combining (\ref{frob}) with (\ref{nullcoord}) we obtain the statement that will be used constantly below, namely that a null congruence whose tangent field is a gradient is twist free, while a twist free null congruence is locally proportional to the gradient of a null function. The comparison of (\ref{grad1}) with (\ref{frob}) shows that the twist is precisely the obstruction to writing the null direction field as the gradient of a null function, or equivalently the obstruction to the existence of a family of null hypersurfaces orthogonal to the congruence.

\subsection{How a metric specifies a congruence}
\label{sec:whichcongruence}

It is often said that expansion, shear and twist are properties of a congruence and not of a metric, and this is correct, so it is worth being precise about it, because the theorem will be stated in a purely metric form and it is essential to see why the two points of view are compatible. A metric by itself does not single out a congruence, since $\theta$, $\sigma$ and $\omega$ are defined only after a nowhere vanishing $k^a$ has been selected. In practice, however, a congruence is singled out by the metric in one of a small number of canonical ways, and in every case that matters here the choice is forced.

The first route is a null coordinate already present in the ansatz, which is the situation for almost all wave like metrics. For the Brinkmann metric \cite{brin1925}
\be
\label{ppwave}
\dd s^2=-2\,\dd u\,\dd v+H(u,x,y)\,\dd u^2+\dd x^2+\dd y^2
\en
the inverse metric has $g^{uu}=0$, so $u$ solves (\ref{eikonal}) and is a null coordinate. The associated congruence is generated by $k^a=g^{ab}\p_bu=-\p_v$, where the sign follows from $g^{uv}=-1$, and since $\p_vg_{\mu\nu}=0$ one finds
\be
\nabla_ak_b=\tfrac12\big(\p_ag_{bv}+\p_bg_{av}-\p_vg_{ab}\big)=0 ,
\en
so that $k$ is covariantly constant, a CCNV in the classification of Jordan, Ehlers and Sachs \cite{JES1960}. Consequently $\theta=\sigma=\omega=0$ for this congruence, and $k_a=(\dd u)_a$ is a gradient, in agreement with (\ref{grad1}). We record for later use that for (\ref{ppwave}) the covariant statement that $u$ is a null coordinate is equivalent to the existence of a covariantly constant null vector.

The second route is a Killing vector. If the metric possesses a nowhere vanishing null Killing vector $\xi$, the congruence of its integral curves is canonical, and for (\ref{ppwave}) the choice $\xi=\p_v$ coincides with the first route. This is also the relevant choice in Misner space and in the Misner type solutions of Sec.~\ref{sec:ahmed}. The third route is a principal null direction. For algebraically special spacetimes the metric singles out repeated principal null directions of the Weyl tensor, and by the Goldberg--Sachs theorem \cite{GS1962} in four-dimensional vacuum gravity, a geodesic shear-free null congruence exists if and only if the Weyl tensor is algebraically special. In the non-conformally-flat case the congruence follows a repeated principal null direction. This is the canonical congruence of Kerr and Schwarzschild, and its optical properties are useful when comparing those backgrounds. A fourth route is a matter congruence, since for a perfect fluid with four velocity $u^a$ the vorticity of $u$ is a physical observable, and this is the relevant choice in G\"odel's universe, where the dust congruence is geodesic, expansion free and shear free but has non-vanishing vorticity. The last route is the gradient of a scalar, which is the choice made by the theorem, because once a function satisfying (\ref{eikonal}) has been found the congruence of $\nabla u$ is canonical and, by (\ref{grad1}), twist free.

The important point is that these routes can disagree, and that when they do only one of them is relevant for the question of whether the spacetime can contain a closed timelike curve. The Schwarzschild exterior, for instance, is expanding in its principal null congruence, with $\theta=2/r$ in Eddington--Finkelstein coordinates, shear free and twist free, and it also admits a global null coordinate. Kerr, by contrast, has a principal null congruence that is still geodesic and shear free but twisting, since the complex expansion of the outgoing principal null direction in Boyer--Lindquist coordinates is
\be
\label{kerrrho}
\rho=-\frac{1}{r-ia\cos\theta}\qquad\Longrightarrow\qquad \mathrm{Im}\,\rho=-\frac{a\cos\theta}{r^2+a^2\cos^2\theta}\neq0
\en
for $a\neq0$ and $\theta\neq\pi/2$, whereas for Schwarzschild $\rho=-1/r$ is real. The list of routes is therefore a list of tests rather than a source of ambiguity, since the theorem uses the last of them and we diagnose its failure by inspecting whichever canonical congruence is available.

We close this subsection with a remark that removes any residual uneasiness about assigning a congruence to a metric. The hypothesis of the theorem, that is the existence of a global function with null gradient, is not a statement about a chosen congruence. It is the statement that the foliation of $\M$ by the level sets of $u$ is null, equivalently that some null direction field is hypersurface orthogonal, equivalently that the null eikonal equation (\ref{eikonal}) has a global solution. We may test it without ever committing to a congruence, and when it holds the congruence it selects is the twist free one of the last route, for which the Frobenius condition (\ref{frob}) is satisfied by construction.

\subsection{The shape of the hypothesis}
\label{sec:shape}

It is convenient to restate the hypothesis in a form that makes contact with the causality literature. We call property \textbf{(N)} the assertion that there exists $u\in C^1(\M)$ with $g^{ab}\p_au\,\p_bu=0$ and $\p_au\neq0$ everywhere. Since $\nabla u$ is null but not timelike, \textbf{(N)} is not the same as the existence of a time function, and Sec.~\ref{sec:relation} shows that the two are in fact incomparable, which is the source of the additional mileage of the theorem.

\section{The theorem}
\label{sec:thm}

We now state the result and prove it, and we separate the logical content from the applications because the proof is short and its brevity is itself part of the claim. The section has four subsections. We begin with the statement and with the terminology of closed, future directed and piecewise $C^1$ curves that the statement requires. We then give the proof, which runs in three steps and rests on a single algebraic fact about null covectors and timelike vectors, and we follow it with the regularity assumptions and the immediate corollaries, the first of which extends the conclusion from a null gradient to any causal gradient that never vanishes. We end with a set of metrics in which the criterion follows by inspection, namely the Brinkmann waves and the Kundt class, for which Lemma 1 shows that the expansion, the shear and the twist vanish identically, so that a reader who intends to use the theorem can see how little calculation it demands.

\subsection{Statement}

We use the following terminology. A closed timelike curve is a closed, future directed, piecewise $C^1$ curve $\gamma:[0,1]\to\M$ with $\gamma(0)=\gamma(1)$ and $\dot\gamma^a\dot\gamma_a<0$ at every point at which $\dot\gamma$ exists, and a closed null curve is defined analogously with $\dot\gamma^a\dot\gamma_a=0$ and $\dot\gamma\neq0$.

\begin{quote}
\noindent\textit{Theorem 1.} Let $(\M,g)$ be a smooth and connected four dimensional spacetime. Suppose that there exists a function $u\in C^1(\M)$ such that
\be
\label{hyp}
g^{ab}\,\p_au\,\p_bu=0\qquad\text{everywhere on }\M,\qquad\text{and}\qquad \p_au\neq0\qquad\text{everywhere on }\M .
\en
Then $\M$ contains no closed timelike curve.
\end{quote}

By Sec.~\ref{sec:nullcoord} the hypothesis (\ref{hyp}) is equivalent to each of three other statements, namely that the null eikonal equation (\ref{eikonal}) admits a global $C^1$ solution, that there exists a nowhere vanishing null direction field on $\M$ that is hypersurface orthogonal and whose orthogonal null hypersurfaces foliate $\M$ transversally to a global function, and that $\M$ admits a global null foliation whose leaves are the level sets of a single valued function. We refer to (\ref{hyp}) as \textbf{(N)} throughout.

\subsection{Proof}

The proof proceeds in three steps.

In the first step we show that a null covector is never orthogonal to a timelike vector. We set $k_a=\p_au$, so that $k^ak_a=0$ by (\ref{hyp}), and we take $V^a$ timelike, $V^aV_a<0$. Suppose for contradiction that $V^ak_a=0$. Then $k_a$ lies in the orthogonal complement $V^\perp$ of the timelike vector $V$, and since $V$ is timelike this complement is a spacelike three dimensional subspace of the cotangent space, so that every non-zero element $w\in V^\perp$ satisfies $w^aw_a>0$. Choosing a basis of $V^\perp$, the restriction of the metric to it is positive definite, so $k$, being a non-zero element of $V^\perp$ with $k^ak_a=0$, would be a vector of zero norm in a positive definite space, which is impossible. Hence
\be
\label{step1}
V^aV_a<0\quad\Longrightarrow\quad V^ak_a\neq0 .
\en
This step uses only the algebraic fact that the orthogonal complement of a timelike vector is spacelike, that is, only the Lorentzian signature. It uses neither the field equations nor an energy condition, nor compactness, nor completeness, nor any global property other than the existence of $u$.

In the second step we show that the sign of $V^ak_a$ is the same for all future directed timelike $V$. We fix $k$ as above and consider the set
\be
\label{futurecone}
\mathcal{T}^+=\{\,V\in T_p\M:\ V^aV_a<0,\ V\ \text{future directed}\,\},
\en
which is an open convex cone, because if $V,W\in\mathcal{T}^+$ and $\lambda\in[0,1]$ then $\lambda V+(1-\lambda)W$ is future directed and
\be
\big(\lambda V+(1-\lambda)W\big)^2=\lambda^2V^2+(1-\lambda)^2W^2+2\lambda(1-\lambda)V\!\cdot\!W<0 ,
\en
the inequality being the reverse Cauchy--Schwarz inequality for two future directed timelike vectors. The functional $\Phi(V)=V^ak_a$ is linear on $T_p\M$, and by (\ref{step1}) it does not vanish on the convex set $\mathcal{T}^+$. A linear functional that is everywhere non-zero on a connected convex set has constant sign on it, so
\be
\label{step2}
\text{there is }\alpha\in\{+1,-1\}\text{ such that }\ \alpha\,V^ak_a>0\ \ \text{for all }V\in\mathcal{T}^+ ,
\en
and by continuity $\alpha\,V^ak_a\ge0$ for every future directed causal $V$, with equality only when $V\propto k$.

In the third step we show that $u$ is monotone along causal curves and strictly monotone along timelike ones. We take a future directed causal curve $\gamma:[0,1]\to\M$ and set $f(s)=u(\gamma(s))$. On each smooth piece of $\gamma$ the function $f'$ is continuous and, by (\ref{step2}), nowhere zero, so it has constant sign there. At a corner the one sided tangents are limits of timelike vectors and are therefore future directed causal vectors at the same point, so by (\ref{step2}) the two one sided limits of $f'$ are non-negative, neither piece can carry a negative $f'$, and $f'$ cannot change sign across the corner. Hence
\be
\label{step3}
f'(s)=\dot\gamma^a\p_au=\dot\gamma^ak_a \ \begin{cases} \ge 0 ,\\[2pt] >0 & \text{if }\dot\gamma \text{ is timelike and } \alpha=+1 ,\\[2pt] <0 & \text{if }\dot\gamma \text{ is timelike and } \alpha=-1 .\end{cases}
\en
Consequently $f$ is weakly monotone on $[0,1]$ and strictly monotone on every subinterval on which $\gamma$ is timelike. A closed timelike curve contains such a subinterval by definition, so either $f(1)>f(0)$ or $f(1)<f(0)$, whereas $\gamma(0)=\gamma(1)$ gives $f(1)=f(0)$. This is a contradiction, and therefore no closed timelike curve exists. $\square$

\subsection{Remarks and corollaries}

The proof uses $u\in C^1$ and $\gamma$ piecewise $C^1$. The same conclusion holds for a merely locally Lipschitz $u$, because then $f=u\circ\gamma$ is absolutely continuous and weakly monotone, and is strictly increasing where $\gamma$ is timelike, so that a closed curve again forces $f(1)>f(0)$. For the applications below $C^\infty$ is available.

Null curves are not excluded. If $\gamma$ is null then (\ref{step2}) gives only $f'\ge0$, and $f'=0$ precisely when $\dot\gamma\propto k$, that is when $\gamma$ is tangent to the null direction of the foliation. Such a curve lies inside a single leaf $u=\mathrm{const}$ and may well close. The hypothesis (\ref{hyp}) therefore forbids closed timelike curves while permitting closed null curves, and this is not a defect, since the boundary of a chronology violating region is generated by null geodesics, which are closed in the compactly generated case, and the theorem is exactly the statement that such a curve cannot be deformed into a timelike one.

Only the causal character of $\nabla u$ is used in the proof, and the argument therefore yields a stronger statement, which we record for later use.

\begin{quote}
\noindent\textit{Corollary 1.} If there exists a global $C^1$ function $u$ whose gradient is everywhere causal and nowhere zero, then $\M$ contains no closed timelike curve.
\end{quote}

The timelike case of Corollary 1 is Hawking's time function lemma \cite{hawking1969}. Theorem 1 is therefore an extension of that criterion rather than a weakening of it, since it asserts that the conclusion survives when the time function is replaced by a null coordinate, and Sec.~\ref{sec:relation} shows that the extension is not vacuous.

No field equation and no energy condition is used. The hypothesis (\ref{hyp}) is a pointwise differential condition with a global ingredient that is purely functional, namely the single valuedness of $u$. The theorem therefore makes no assumption about the matter content, about completeness, or about whether $\M$ is globally hyperbolic, and it applies equally to vacuum solutions and to solutions supported by arbitrary matter.

The theorem is one-way. The converse fails, and the simplest witnesses are the closed cosmological models, since a spacetime with a compact Cauchy surface cannot satisfy the hypothesis at all, as we prove in Sec.~\ref{sec:relation}, and yet the closed Friedmann models are stably causal and free of closed timelike curves. It is worth recording in passing that the maximally extended Schwarzschild solution, which one might expect to provide a second witness by virtue of its horizon, does not. In Kruskal coordinates the metric reads $f(-\dd T^2+\dd X^2)+r^2\dd\Omega^2$ with $f>0$, so that $U=T-X$ obeys $g^{ab}\p_aU\,\p_bU=0$ and is a global null coordinate. The failure of a particular chart to be global therefore says nothing about the existence of other global null functions. The fact that the hypothesis is merely sufficient is what makes the theorem usable, since when it applies nothing else needs to be checked.

The foliation is forced and not chosen. No congruence is selected by hand in the proof, the only object that enters being the level set of $u$, and $u$ is fixed by the metric through (\ref{eikonal}). By (\ref{grad1}) and (\ref{frob}) the corresponding null congruence is automatically twist free, with vanishing twist
\be
\label{omega0}
\omega^2=\tfrac12\nabla_{[a}k_{b]}\nabla^ak^b=\tfrac12\nabla_{[a}\nabla_{b]}u\,\nabla^{[a}\nabla^{b]}u=0 ,
\en
because $\nabla_a\nabla_bu$ is symmetric. Equation (\ref{omega0}) is the technical heart of the theorem, since it says that the twist of the congruence orthogonal to the null foliation vanishes identically.

\subsection{The theorem in practice}
\label{sec:practice}

For the Brinkmann metric (\ref{ppwave}) the function $u=t-z$ satisfies $g^{ab}\p_au\,\p_bu=g^{uu}=0$ identically, for every profile $H(u,x,y)$, and $\p_au\neq0$. The theorem then says, with no calculation whatsoever, that a pp-wave spacetime contains no closed timelike curve. We record this as a corollary, and we extend it at once to the whole Kundt class with the help of a lemma.

\begin{quote}
\noindent\textit{Lemma 1.} For the Kundt metric
\be
\label{kundt}
\dd s^2=-2\,\dd u\,\dd v+F(u,v,x,y)\,\dd u^2+2W_i(u,x,y)\,\dd u\,\dd x^i+g_{ij}(u,x,y)\,\dd x^i\dd x^j ,
\en
with $\p_vW_i=\p_vg_{ij}=0$, the vector $k^a=\p_v$ is null and geodesic, its covariant derivative is
\be
\label{kundtderiv}
\nabla_ak_b=\tfrac12\,\p_vg_{ab}=\tfrac12\,(\p_vF)\,k_ak_b ,
\en
and consequently
\be
\label{kundtopt}
\theta=0,\qquad \sigma=0,\qquad \omega=0 ,
\en
for every choice of the rotation one form $W=W_i\dd x^i$ and every profile $F$.
\end{quote}

\noindent\textit{Proof.} The inverse metric of (\ref{kundt}) has $g^{uu}=0$ and $g^{uc}=(0,-1,0,0)$, so that
\be
\Gamma^{u}{}_{ab}=-\tfrac12\big(\p_ag_{bv}+\p_bg_{av}-\p_vg_{ab}\big)=\tfrac12\,\p_vg_{ab} ,
\en
where the last equality uses the fact that $g_{bv}$ is constant, since the rotation enters $g_{ui}=W_i$ and not $g_{iv}$. The lowered components of $k$ are therefore $k_a=g_{av}=(-1,0,0,0)$, which is to say that $k=-\dd u$ exactly, and
\be
\nabla_ak_b=\p_ak_b-\Gamma^{c}{}_{ab}k_c=0+\Gamma^{u}{}_{ab}=\tfrac12\,\p_vg_{ab} .
\en
Since $\p_vg_{ab}=\p_vF\,\delta^{u}_{a}\delta^{u}_{b}$ is symmetric and proportional to $k_ak_b$, it lies entirely in the direction orthogonal to the screen space, so $\omega=0$ and $\sigma=0$. Moreover it is non-zero only for $a=b=u$, so
\be
\theta=g^{ab}\nabla_ak_b=\tfrac12 g^{uu}\,\p_vF=0 ,
\en
because $g^{uu}=0$. Hence the lemma follows. $\square$

Two remarks on Lemma 1 are in order, because the second corrects a natural but mistaken expectation. The first is that the only non-trivial component of $\nabla k$ is the term proportional to $\p_vF$, which is the recurrence term familiar from the classification of Kundt waves \cite{stephani2003,griffiths2009} and which vanishes for the pp-wave subclass. The hypothesis $\p_vW_i=0$ of the lemma is what reduces $\nabla_ak_b$ to that single term. If $W$ is allowed to depend on $v$, as the most general Kundt metrics permit, then $\nabla_ak_b=\tfrac12\p_vg_{ab}$ acquires in addition a symmetric piece with mixed $(u,i)$ indices, so the identity $\nabla_ak_b\propto k_ak_b$ fails, but the conclusion $\theta=\sigma=\omega=0$ survives, because $g^{ui}=0$ still annihilates the trace and the mixed piece is symmetric. Corollary 2, which rests only on $g^{uu}=0$, is therefore unaffected. The second is that the rotation one form $W$ does not appear in (\ref{kundtderiv}) at all. The one form $W$ is not a twist of the congruence, since the cross term in (\ref{kundt}) is $2W_i\,\dd u\,\dd x^i$ and therefore contributes to $g_{ui}$ rather than to $g_{iv}$, and its curl enters instead the field equations through the Ricci component $R_{ui}$. Physically $W$ is a gravito-magnetic potential that endows the wave with angular momentum while leaving the null foliation intact, and the congruence remains hypersurface orthogonal however large $W$ becomes.

\begin{quote}
\noindent\textit{Corollary 2.} No Brinkmann spacetime of the form (\ref{ppwave}) and no Kundt spacetime of the form (\ref{kundt}), with any profile $H$ or $F$ and any rotation one form $W$, contains a closed timelike curve on the domain where $u$ is a global coordinate. The same statement holds for plane waves, sandwich waves, impulsive and shock waves, and for the gyratonic waves obtained by endowing (\ref{ppwave}) with a non-zero $W_i$.
\end{quote}

Corollary 2 follows at once from Lemma 1 and Theorem 1, since $g^{uu}=0$ for the whole class. We emphasise that the qualification on the rotation one form that one might have expected is absent, and that this is a consequence of the lemma rather than an assumption.

\section{Sharpness and the two escape routes}
\label{sec:sharp}

A sufficient condition is useful only if we know how it can fail, and here we show that within the class of spacetimes considered in this paper the hypothesis fails in exactly two ways, that each failure is a sharp geometric property, and that both occur in known chronology violating spacetimes, which turns Theorem 1 from a test into a classification. We divide the argument into five subsections. We first split the hypothesis into its two independent requirements, twist freedom and the global single valuedness of the null coordinate, since each escape route corresponds to the failure of one of them. We then treat the first route, in which the congruence twists and no null leaf exists, and we prove that a twisting congruence is never proportional to a gradient. We next treat the second route, in which the congruence is twist free but the null coordinate fails to be single valued and the leaves close up, and we derive the condition under which the compactified direction turns timelike. We then examine traversable wormholes, which are the other standard setting for time machines and which turn out to add a topology but not a third route, since a static wormhole satisfies the hypothesis with an explicit null coordinate while a wormhole time machine fails it by exactly the monodromy that the second route describes. We close with the null cylinder, which shows that the hypothesis does not follow from stable causality, so that we cannot relax it into that condition.

\subsection{The structure of the hypothesis}

\begin{figure}
\centering
\includegraphics[width=0.99\textwidth]{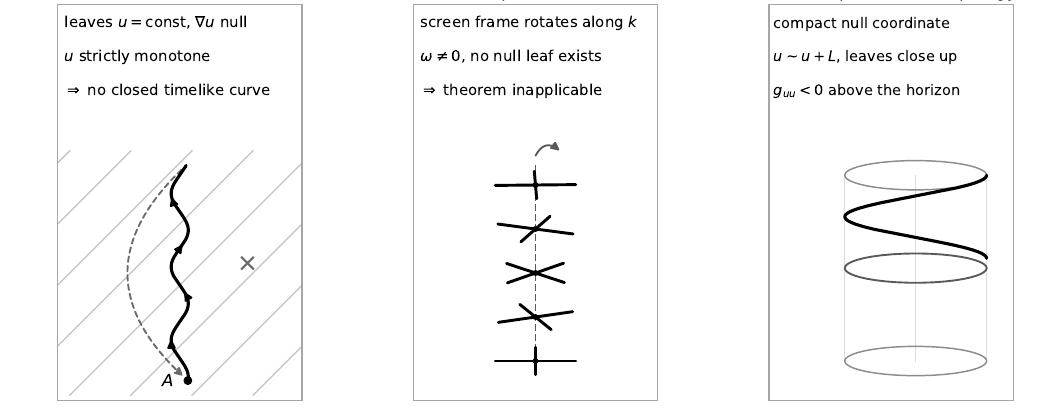}
\caption{Panel (a) shows the obstruction. The level sets $u=\mathrm{const}$ of a null coordinate foliate the spacetime, and a timelike curve crosses every leaf transversally, so it cannot close. Panel (b) shows the first escape route, in which the screen frame rotates along the congruence so that no family of null leaves exists and $\omega\neq0$. Panel (c) shows the second escape route, in which the null coordinate is compactified, so that a null leaf still exists locally but the leaves close up and a timelike curve of the universal cover runs from $u$ to $u+nL$.}
\label{fig:nogo}
\end{figure}

The hypothesis (\ref{hyp}) is the conjunction of two independent requirements. The first, which we call $(\alpha)$, is twist freedom, meaning that the null direction field under consideration must be hypersurface orthogonal so that a null leaf passes through every point. The second, which we call $(\beta)$, is global single valuedness, meaning that the null coordinate built on those leaves must be a genuine function on $\M$ and must not have a period. If $(\alpha)$ fails then no null foliation exists at all and the theorem is silent, while if $(\beta)$ fails then a null foliation exists locally, and on the universal cover, but the leaves close up and the theorem is again silent. We should add that a third logical possibility exists in principle, namely that no global nowhere vanishing null direction field exists at all, since the bundle of null lines over $\M$ is a two sphere bundle that need not admit a section. This possibility never arises for the spacetimes considered here, all of which are parallelisable patches of $\R^4$, and we mention it only to avoid overstating the dichotomy. The two routes are illustrated in Fig.~\ref{fig:nogo}.

\subsection{Escape route one, twist}
\label{sec:esc1}

\begin{quote}
\noindent\textit{Proposition 1.} Let $k^a$ be a nowhere vanishing null geodesic congruence on $\M$ with twisting $\omega^2>0$. Then $k$ is not proportional to the gradient of any function and there is no family of null hypersurfaces orthogonal to $k$. Conversely, if $\M$ admits a nowhere vanishing twist free null direction field whose null foliation has leaf space diffeomorphic to $\R$, then $\M$ admits a global null coordinate and, by Theorem 1, no closed timelike curve.
\end{quote}

\noindent\textit{Proof.} The first statement is (\ref{grad1}) together with (\ref{frob}). For the converse we use the Frobenius form of twist freedom, which is the hypothesis actually at hand. Since $k\wedge\dd k=0$ the distribution $k^\perp$ is integrable, so that $\M$ is foliated by null hypersurfaces. The one form $k$ annihilates the distribution and is nowhere zero, so on each chart $k=f\,\dd u$ with $f\neq0$, and on overlaps the local functions differ by reparametrisations rather than by constants. If the leaf space is $\R$, these functions patch into a global $u$. The closedness of $k$, that is $\nabla_{[a}k_{b]}=0$, would also make $k^\perp$ integrable, but it is sufficient rather than necessary and the proof does not need it. $\square$

The hypothesis on the leaf space is not a technicality, as Misner space of Sec.~\ref{sec:ahmed} shows. There the field $k=-\p_t$ is globally defined and null, with $k_a=\dd\psi$, so it is twist free on the whole manifold, and its null leaves $\psi=\mathrm{const}$ have a leaf space which is a circle, a Hausdorff one dimensional manifold. Should the weaker hypothesis suffice, the proposition would conclude that Misner space has a global null coordinate and no closed timelike curve, whereas in fact it has neither, since $\psi$ is not a function on $\M$ and the circles of constant $t>0$ are closed timelike curves. The leaf space must therefore be $\R$, and the reason is visible in the proof, since a circle valued null coordinate cannot be exponentiated into a real function. Simple connectivity of $\M$ is not a substitute either, because the Reeb foliation of $S^3$ shows that a simply connected manifold can carry a codimension one foliation whose leaf space fails to be Hausdorff. In all the applications below the global null coordinate is exhibited explicitly, so these distinctions are moot there.

The physical reading of Proposition 1 is that twist is a resource for chronology violation. It is the geometric quantity that allows a null congruence to spin in such a way that its integral curves cannot be simultaneously orthogonal to a family of surfaces, so that a monotone null function is not available. Two features of the criterion deserve emphasis in the light of Lemma 1. The first is that the Kundt class is not an example. Since the rotation one form $W$ of (\ref{kundt}) does not enter $\nabla k$ at all, the Kundt congruence is twist free for every $W$, and the intuition that a rotating wave should produce a twisting congruence is wrong. The second is that the twist route therefore requires a congruence that is neither non-expanding nor shear free, which makes the route considerably less accessible than one might have supposed. The minimal known example is the principal null congruence of Kerr, and its twist is given explicitly in (\ref{kerrrho}).

The classical chronology violating solutions confirm the criterion. Kerr's principal null congruence is geodesic and shear free but twisting, with $\mathrm{Im}\,\rho\neq0$ whenever $a\neq0$ and $\theta\neq\pi/2$, and Kerr contains closed timelike curves near the ring singularity, whereas the Schwarzschild congruence has $\rho=-1/r$ real, is twist free, and admits the global null coordinate $u=t-r_*$ on the exterior. G\"odel's dust is geodesic, expansion free and shear free but has non-vanishing vorticity, and it is totally vicious. A further candidate, which has not been analysed from this point of view, is Hauser's twisting type N solution \cite{hauser1974}, the only vacuum solution of that type known to us, whose repeated principal null congruence is expanding and twisting and which therefore falls outside the reach of Theorem 1. Whether it contains closed timelike curves is, to the best of our knowledge, an open question.

\subsection{Escape route two, topology}
\label{sec:esc2}

The second escape route is topological, and it is the mechanism behind essentially all exact chronology violating solutions constructed from regular initial data. Suppose that a null coordinate $u$ fails to be single valued because it is identified,
\be
\label{compact}
u\sim u+L,\qquad L>0 ,
\en
with the other coordinates unchanged. Then $\M$ is a quotient of the spacetime on which $u$ was defined rather than that spacetime itself, and $u$ is not a function on $\M$.

\begin{quote}
\noindent\textit{Corollary 3.} Let $(\tilde\M,\tilde g)$ admit a global null coordinate $u$, and let $\M=\tilde\M/\Gamma$ with $\Gamma\cong\Z$ acting as in (\ref{compact}). Then every closed timelike curve of $\M$ lifts to a timelike curve of $\tilde\M$ running from a point $p$ to its translate $\gamma(p)$, with $\gamma\in\Gamma$ and $\gamma\neq\mathrm{id}$, along which $u$ increases by the period $nL\neq0$.
\end{quote}

\noindent\textit{Proof.} Since $\tilde\M\to\M$ is a local isometry and the lift of any timelike curve has strictly monotone $u$ by Theorem 1, the lift of a closed timelike curve is timelike and connects distinct translates. The curve is therefore closed in $\M$ but not in $\tilde\M$. $\square$

Corollary 3 says that in this setting closed timelike curves are entirely topological, since they exist in the quotient if and only if a timelike curve in the cover separates a point from its translate. It also explains why no energy condition need be violated by such a construction, the local geometry being everywhere perfectly regular. For the metrics used in practice the condition takes an even simpler form. We write the metric in coordinates adapted to the compact null direction as
\be
\label{misnerform}
\dd s^2=2g_{u\mu}\,\dd u\,\dd x^\mu+g_{\mu\nu}\,\dd x^\mu\dd x^\nu ,
\en
where the block $g_{\mu\nu}$ carries no $v^2$ term and no $vx^i$ cross terms, that is $g_{vv}=g_{vi}=0$, so that $g^{uu}=0$ holds identically and $u$ is a null coordinate on the cover, and we consider the curve obtained by fixing $x^\mu$ and letting $u$ wind once around the circle. This curve is closed by construction, and it is timelike if and only if $g_{uu}<0$ at every point of it.

\begin{quote}
\noindent\textit{Corollary 4.} For a metric of the form (\ref{misnerform}) with the identification (\ref{compact}), the spacetime contains a closed timelike curve whenever the compact direction $\p_u$ is timelike at every point of the circle obtained by holding the remaining coordinates fixed, that is whenever $g_{uu}<0$ along that whole circle. In the Misner type examples of Sec.~\ref{sec:ahmed} the hypersurface $g_{uu}=0$ is the chronology horizon, on which the corresponding curve is a closed null curve.
\end{quote}

We emphasise the words along that whole circle, because the weaker condition that $g_{uu}<0$ at some point is not sufficient, and the distinction matters for Ori's metric, in which the profile depends on the compact coordinate. We shall see in Sec.~\ref{sec:ahmed} that Corollary 4 reproduces, in one line each, the chronology conditions of Misner space, of Ori's time machine and of the Misner type solutions of Ahmed. The two corollaries are asymmetric, since Corollary 3 is an exact statement about where closed curves can come from whereas Corollary 4 gives a sufficient condition that is operative in the Misner type examples analysed below. The converse of Corollary 4 fails, and not marginally. A quotient of the form (\ref{misnerform}) with $g_{uu}>0$ everywhere can contain a closed timelike curve along which all coordinates vary, because the closed curve need not follow the compact generator. The analytic condition remains the existence of a timelike curve in the cover from $(0,x^\mu)$ to $(L,x^\mu)$, and the plane-wave quotients of Hubeny, Rangamani and Ross \cite{HRR2004} realise it explicitly. We return to this point in Sec.~\ref{sec:recipe}.

\subsection{Wormholes, a third topology and not a third route}
\label{sec:wormhole}

Traversable wormholes are the other standard setting for time machine constructions, and it is worth being explicit about where they sit, because they are neither wave like nor compact and a reader may suspect that they constitute a third escape route. They do not. The Morris--Thorne wormhole \cite{mor1988} reads, in the proper length $l$ along the throat,
\be
\label{mt}
\dd s^2=-e^{2\Phi(l)}\,\dd t^2+\dd l^2+r(l)^2\,\dd\Omega^2 ,
\en
with $l\in\R$, so that the two asymptotically flat regions correspond to $l\to\pm\infty$ and the throat sits at the minimum of $r$, where $r_{,l}=0$. The only condition that concerns us is that $\Phi$ be finite, since then $g_{tt}=-e^{2\Phi}$ never vanishes, no horizon forms and $t$ is a global time function. The question is whether such a spacetime also admits a global null coordinate, and the answer is that it does, with
\be
\label{mtnull}
u=t+F(l),\qquad F_{,l}=e^{-\Phi(l)} ,
\en
because then
\be
\label{mtnullcheck}
g^{ab}\p_au\,\p_bu=g^{tt}+g^{ll}(F_{,l})^2=-e^{-2\Phi}+e^{-2\Phi}=0 ,
\en
while $\p_au=(1,F_{,l},0,0)$ never vanishes. Since $F$ is the integral of a continuous function of $l$ it is single valued on the whole wormhole, the two sides included, and no identification is involved. Hence \textbf{(N)} holds, Theorem 1 applies, and a static traversable wormhole contains no closed timelike curve, in agreement with the fact that it is stably causal. No field equation has entered, so the exotic matter that supports the throat is irrelevant to the argument.

The Ellis wormhole \cite{ellis1973} is the case $\Phi=0$ with $r^2=l^2+b^2$, for which (\ref{mtnull}) reduces to the statement that $u=t+l$ is a null coordinate, which one verifies at sight. The general static wormhole, in which the spatial slice carries an arbitrary metric $h_{ij}$ rather than the spherically symmetric one of (\ref{mt}), admits a null coordinate of the same form $u=t+F(x^i)$ whenever the eikonal equation
\be
\label{mthj}
h^{ij}\p_iF\,\p_jF=e^{-2\Phi}
\en
has a global solution on the spatial slice. This is a Hamilton--Jacobi equation with a convex Hamiltonian, so a global Lipschitz solution exists whenever $e^{-\Phi}$ is bounded, and the Lipschitz form of Theorem 1 recorded in Sec.~\ref{sec:thm} then suffices. In this sense the criterion covers the whole static wormhole class and not merely the symmetric examples.

Two features of (\ref{mtnull}) deserve comment. The first is that the congruence is not of Kundt type, so the wormhole illustrates the irrelevance of expansion in a way that is quite independent of the Robinson--Trautman case of Sec.~\ref{sec:rt}. For $k=\nabla u$ one finds
\be
\label{mtopt}
\theta=\frac{2e^{-\Phi}r_{,l}}{r},\qquad \sigma=0,\qquad \omega=0 ,
\en
so that the congruence focuses on the way into the throat and defocuses on the way out, with $\theta$ changing sign where $r_{,l}=0$. The expansion is large and takes both signs, the shear vanishes, the twist vanishes by construction, and the theorem is indifferent to all of it, since all it requires is that $\nabla u$ be null. The second is that the null coordinate (\ref{mtnull}) is exactly what a wormhole time machine has to destroy. The construction of Morris, Thorne and Yurtsever \cite{mty1988} gives the two mouths different clock rates, by accelerating one of them or by placing it in a gravitational potential, so that a voyage through the throat arrives at the other mouth at a time shifted by an amount $\Delta T$, and a loop through the throat and back outside then closes and becomes timelike once $\Delta T$ exceeds the external travel time. Such a spacetime contains closed timelike curves and therefore, by the contrapositive of Theorem 1, does not satisfy \textbf{(N)}. The failure is not twist, since the local geometry is that of the twist free congruence (\ref{mtnull}); what fails is single valuedness, because the gluing that produces the time shift destroys the global time coordinate and with it the null coordinate (\ref{mtnull}), leaving behind a monodromy equal to the accumulated time difference.

Wormholes therefore add a topology but not a route. In Misner space the closed loop is a circle of the null coordinate itself, whereas in a wormhole time machine it is a loop through the throat whose holonomy in the null coordinate is the accumulated time difference between the mouths. Both are instances of $(\beta)$ failing, and the classification of Sec.~\ref{sec:sharp} applies unchanged. What the wormhole does add is a caution about language, since the phrase time machine suggests a device placed inside a spacetime, whereas what the theorem sees is a closed loop around which no monotone function can exist.

\subsection{A counterexample to stable causality}
\label{sec:counterexample}

We now show that the hypothesis cannot be relaxed, and in doing so we settle the logical question raised in Sec.~\ref{sec:intro}, since a global null coordinate does not imply stable causality. We consider the null cylinder
\be
\label{nullcyl}
\dd s^2=-2\,\dd u\,\dd v+\dd x^2+\dd y^2 ,\qquad (u,v,x,y)\in S^1\times\R\times\R^2 ,
\en
that is, flat space with one null direction compactified. Independently of the compactification of $u$, the coordinate $v$ is a global null coordinate, because $g^{vv}=0$ for the metric (\ref{nullcyl}) with $v\in\R$. The hypothesis (\ref{hyp}) is therefore satisfied with $u$ replaced by $v$, and Theorem 1 applies, so the null cylinder contains no closed timelike curve. This is easy to confirm directly, since a curve winding around the $u$ circle must have $\dot v$ of constant sign if it is to remain causal, so $v$ is monotone and the curve cannot close unless $\dot v\equiv0$, in which case it is null rather than timelike.

On the other hand the null cylinder is not stably causal. It contains closed null curves, namely the curves of constant $(v,x,y)$ winding once around the $u$ circle, and along them $u$ is monotone while $v$ is constant, exactly as Theorem 1 requires. The existence of such curves forbids a time function, since a time function would have to increase strictly around a closed curve. Therefore \textbf{(N)} does not imply \textbf{(T)}, and together with the closed Friedmann models, which are stably causal and which by the lemma proved in Sec.~\ref{sec:relation} admit no global null coordinate, this shows that \textbf{(N)} and \textbf{(T)} are logically independent. Neither implies the other, and both imply the absence of closed timelike curves.

\section{Relation to earlier criteria and what is new}
\label{sec:relation}

The theorem would be of little interest if it merely repeated a known criterion in unfamiliar words, so we now place it among the three families of results recalled in Sec.~\ref{sec:intro} and say precisely what it adds and what it does not. We make the comparison in five subsections. We begin with a logical map of three properties, the chronology condition, stable causality and the hypothesis \textbf{(N)}, and we record which implications hold and which fail, together with the lemma that a compact Cauchy surface excludes \textbf{(N)}. We then compare the criterion with the time function theorems, with the energy condition theorems and with the topological theorems in turn, and in each case we identify the spacetimes on which the two verdicts differ. We end with an explicit account of what is new, which we state as narrowly as the evidence allows, because the observation behind the proof is classical and the novelty lies in its formulation rather than in its difficulty.

\subsection{A logical map}

We fix notation for three properties. Property \textbf{(C)} is the chronology condition, that $\M$ contains no closed timelike curve. Property \textbf{(T)} is stable causality. Property \textbf{(N)} is the hypothesis (\ref{hyp}), that a global $C^1$ function with everywhere null and nowhere zero gradient exists. The two implications used here are
\be
\label{ladder}
\textbf{(T)}\ \Longrightarrow\ \textbf{(C)},\qquad
\textbf{(N)}\ \Longrightarrow\ \textbf{(C)}.
\en
where the first follows from the existence of a time function \cite{hawking1969,bernal2005} and the second is Theorem 1 of this paper.

The classical results of Tipler \cite{tipler1976} and Hawking \cite{hawking1992} address the formation of causality violation under additional geometric and energy hypotheses.

As shown in Sec.~\ref{sec:counterexample}, the null cylinder satisfies \textbf{(N)} and fails \textbf{(T)}, so \textbf{(N)} does not imply \textbf{(T)}. The separation in the other direction is less obvious than it appears, because the maximally extended Schwarzschild solution, which is the first candidate that comes to mind, satisfies \textbf{(N)} through the Kruskal null coordinate $U=T-X$ noted in Sec.~\ref{sec:thm}. The natural witnesses are closed universes, and the reason is a general observation which we record as a lemma.

\begin{quote}
\noindent\textit{Lemma 2.} A spacetime with a compact Cauchy surface does not satisfy \textbf{(N)}.
\end{quote}

\noindent\textit{Proof.} Let $\Sigma$ be a compact Cauchy surface and $u$ a $C^1$ function obeying (\ref{hyp}). Since $\Sigma$ is compact, $u|_\Sigma$ attains its maximum at some $p\in\Sigma$, and the derivative of $u$ along every direction tangent to $\Sigma$ vanishes there, so $\nabla u$ is orthogonal to $T_p\Sigma$. That orthogonal complement is spanned by the unit normal $n$, which is timelike, so $\nabla u=\lambda n$ and $|\nabla u|^2=-\lambda^2$. The gradient is null by hypothesis, so $\lambda=0$, which gives $\nabla u=0$ at $p$ and contradicts $\p_au\neq0$. Hence no such $u$ exists. $\square$

The closed Friedmann models, the Einstein static universe and the closed slicing of de Sitter space each possess a compact Cauchy surface and are globally hyperbolic, and they therefore satisfy \textbf{(T)} while failing \textbf{(N)}. Combining this with the null cylinder, we conclude that \textbf{(N)} and \textbf{(T)} are incomparable, neither implying the other, both implying the absence of closed timelike curves.

Minkowski space illustrates that \textbf{(N)} can hold even when the null generic condition fails identically. The proof of Theorem 1 requires neither null geodesic completeness nor any energy or generic condition.

In the language of the causal ladder, with the chain of implications from non-totally-vicious through chronological, causal, distinguishing, strongly causal, stably causal, causally continuous and causally simple to globally hyperbolic, Theorem 1 establishes the chronological rung, and it does so more strongly than the definition requires, since it also forbids closed causal curves that are not everywhere tangent to the null foliation. Hawking's theorem establishes the strictly higher stably causal rung. The virtue of the lower rung is that it follows from the existence of a global solution of the null eikonal equation, and the price is that the converse fails.

\subsection{Comparison with the time function criteria}
\label{sec:vstimefunction}

The time function criterion is the closest relative of Theorem 1, so it is worth separating what the two share from what divides them. The shared part is the principle of monotonicity, and Corollary 1 is in fact the timelike case of the classical statement, since the proof of Sec.~\ref{sec:thm} runs unchanged when the gradient is everywhere timelike rather than null. What Theorem 1 adds is a wider domain rather than a stronger conclusion. Because the argument uses only that a null covector is never orthogonal to a timelike vector, it survives the replacement of a clock by a light-sheet, and the hypothesis can then be checked from the null eikonal equation alone, without first constructing the smooth time function that the classical criterion presupposes \cite{geroch1970,bernal2005,bernal2006}.

The extension matters because stable causality is a strong hypothesis. It fails in the null cylinder, in every spacetime containing closed null curves, and in every spacetime whose chronology horizon is compactly generated, and these are precisely the spacetimes in which one is interested when asking whether closed timelike curves exist. In such cases the classical criterion says nothing at all, whereas \textbf{(N)} may be perfectly satisfied, as it is in the null cylinder and, by Lemma 1, in every Kundt and every Robinson--Trautman spacetime.

Conversely we should be explicit about what the theorem does not give. It does not characterise the chronology condition, since it is sufficient and the converse fails. It does not imply stable causality, as the null cylinder shows. The theorem gives no energy or completeness condition and cannot be used to conclude anything about geodesic completeness or about the fate of an observer at a chronology horizon. What it does give is a sufficient condition that is tested by solving one scalar equation, together with a diagnosis, through the two escape routes, of why a given construction succeeds or fails.

\subsection{Comparison with the energy condition criteria}
\label{sec:vsenergy}

The theorems collected in this family answer a different question, since they constrain what matter must do if chronology is to fail rather than decide whether it does. Tipler \cite{tipler1976} obtained restrictions on causality violation in asymptotically flat spacetimes, and Hawking \cite{hawking1992} showed that formation from a noncompact initial surface, under his horizon hypotheses, requires a violation of the averaged weak energy condition on the Cauchy horizon. Kay, Radzikowski and Wald \cite{kay1997} carried the obstruction to the quantum level, showing that the usual renormalised stress tensor of a free scalar field in an initially Hadamard state is ill-defined or singular at the base points of a compactly generated Cauchy horizon, while Krasnikov \cite{krasnikov2002} proved a result on maximal extensions that contain no closed causal curves outside the chronological past of the original spacetime. Read together, these results say that chronology violation leaves a trace in the stress tensor, but they do not say whether a given metric contains a closed timelike curve, and Theorem 1 addresses precisely that question.

A small technical remark makes the gap explicit and is worth recording. Tipler's theorem requires, besides the energy condition and completeness, the null generic condition, which fails identically in any flat spacetime. The null cylinder is precisely a complete, flat, averaged-null-energy-condition-respecting spacetime containing closed null geodesics, so the energy condition route is structurally blind to it, whereas the kinematic route of Theorem 1 has no difficulty with it, since $v$ is a global null coordinate and closed timelike curves are excluded irrespective of the closed null geodesics. This is a concrete instance of the complementarity just described. We note in passing that the generic condition can also fail in a merely Ricci flat spacetime, as it does for Minkowski space, and that it can hold in a Ricci flat one, as it does for a generic pp-wave, so flatness and Ricci flatness play different roles here.

\subsection{Comparison with the topological criteria}
\label{sec:vstopological}

Geroch \cite{geroch1967} and Tipler \cite{tipler1976} studied the relation between topology and causality, and Visser \cite{visser2002} reviewed classical and semiclassical aspects of chronology protection. A future Cauchy horizon is compactly generated if its null generators, followed to the past, enter and remain in a common compact set \cite{hawking1992,kay1997}. In Misner space the horizon generators close under the periodic identification. Corollary 3 gives a criterion for lifting closed curves in a quotient.

\subsection{What is new}
\label{sec:newness}

We summarise as honestly as we can the relation of the present work to the existing literature. The elementary observation that a monotone function obstructs closed curves is classical, and the special case of Corollary 1 with a timelike gradient is Hawking's argument, while the observation that a pp-wave, having a global null coordinate, contains no closed timelike curve is presumably familiar to workers in that field, if only implicitly. What we believe is new can be stated in four strands.

The first strand is that the hypothesis is stated in the sharpest usable form, and that its logical relation to stable causality is settled. The two are incomparable rather than ordered, and the criterion therefore applies to the class of spacetimes in which the classical theorem fails, which is exactly the class containing the known chronology violating solutions. The second strand is that the theorem is sharp and that its failure modes classify the physics, since Proposition 1 and Corollary 3 show that there are exactly two ways to evade it within this class, turning a one line lemma into a decision procedure and explaining in a unified way why the known exact solutions rely on an identification of a null direction. The third strand is that the criterion is decoupled both from the choice of congruence and from the field equations, so that it makes no reference to matter, to energy conditions, to asymptotic structure or to the cosmological constant, and so that it resolves the perennial question of which congruence the words expanding, shearing and twisting refer to. The fourth strand, which we regard as the strongest, is that the criterion produces new no-go statements. Applying it yields the absence of closed timelike curves in the whole Kundt class, for arbitrary cosmological constant and arbitrary rotation one form, and in the whole Robinson--Trautman class, on any region where the null coordinate is a globally single valued function, results which we have not found in the literature and which settle the status of the gyratonic waves.

\section{Application to Misner space, Ori's time machine and the Ahmed series}
\label{sec:ahmed}

In this section we show that the chronology conditions of the standard Misner type constructions, and of the recent series of four dimensional solutions with and without a cosmological constant, follow at once from Corollary 4. We also identify the common mechanism, which is always the compactification of a null coordinate and never the expansion, the shear, the twist or the cosmological constant. We proceed in four subsections. We first rederive the Misner mechanism itself, since it is the prototype and the only case in which we can display every step. We then treat Ori's time machine, where the compact direction is toroidal and where we perform the identification on a regular and asymptotically flat background, and we recover his chronology condition without solving any field equation. We next turn to the Ahmed series, where we need a structural lemma on the cofactor of the metric before the same reasoning applies, which is why the calculation runs a little longer there. We close with the recipe that the three examples share, so that the reader can generate further chronology violating solutions and can see at the same time why none of them can escape the theorem.

\subsection{The Misner mechanism}
\label{sec:misner}

Misner space \cite{misner1967} is the simplest chronology violating spacetime. In the form used in the modern literature it reads
\be
\label{misner}
\dd s^2=-2\,\dd t\,\dd\psi-t\,\dd\psi^2+\dd x^2+\dd y^2,\qquad \psi\sim\psi+\psi_0 ,
\en
the metric is flat, and it is obtained from two dimensional Minkowski space by identifying a boost, or equivalently by closing up the boost orbits of Minkowski space. The universal cover of (\ref{misner}) is the wedge of two dimensional Minkowski space on which the boost acts freely, not the whole space, since a boost orbit is a hyperbola and the quotient discards the other wedges.

We apply the theorem. The inverse of the $(t,\psi)$ block is
\be
\label{misnerblock}
g_{ab}=\ba{cc}0&-1\\-1&-t\ea\Longrightarrow g^{ab}=\ba{cc}t&-1\\-1&0\ea ,
\en
so $g^{\psi\psi}=0$ and $\psi$ is a null coordinate, with $k_a=\p_a\psi$ and $k^bk_b=g^{\psi\psi}=0$. Unrolling $\psi$ gives a cover which is a region of two dimensional Minkowski space, flat and free of closed timelike curves, and on it the coordinate $\psi$ is globally single valued, so \textbf{(N)} holds and Theorem 1 applies. This is of course obvious, since the cover is a wedge of Minkowski space, but it is instructive to see the theorem reproduce it and to see where the closed curves of Misner space itself come from.

\begin{figure}
\centering
\includegraphics[width=0.86\textwidth]{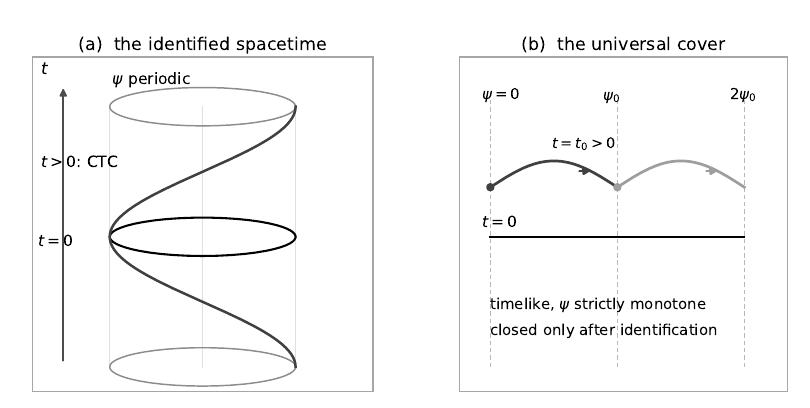}
\caption{The Misner mechanism. The left panel shows the $(t,\psi)$ cylinder, with $t$ vertical and $\psi$ periodic, on which the region $t>0$ is chronology violating and the circle $t=0$ is the chronology horizon. The right panel shows the universal cover, on which $\psi$ ranges over the real line and a timelike curve runs from $\psi=0$ to $\psi=\psi_0$ with $\psi$ strictly monotone, the same curve being drawn again in lighter grey on the next period. It closes only after the identification.}
\label{fig:misner}
\end{figure}

They come from the identification (\ref{compact}). Since $\psi$ is identified with period $\psi_0$, Corollary 3 applies, and the curves of $\M$ winding around the $\psi$ circle lift to timelike curves of Minkowski space joining a point to its boost image. Corollary 4 then gives the condition directly, since the curve of constant $(t,x,y)$ winding once in $\psi$ is timelike if and only if
\be
\label{misnerctc}
g_{\psi\psi}=-t<0\qquad\Longleftrightarrow\qquad t>0 .
\en
The chronology horizon is therefore the null hypersurface $t=0$, which is null because $g^{tt}=t$ vanishes there, and the chronology violating region is the half space $t>0$, in agreement with the standard analysis \cite{misner1967,hawkingellis}. We note how much has been bypassed, since no analysis of the boost identification, no construction of the covering map and no use of the flatness of the metric is required. The only input is the one line observation that $\dd\psi$ is null and $\psi$ is compact. Fig.~\ref{fig:misner} shows the two pictures side by side.

\subsection{Ori's time machine}
\label{sec:ori}

Ori's 2005 time-machine model \cite{ori2005} contains a compact toroidal vacuum core in a regular, asymptotically flat spacetime. Ori's later vacuum--dust construction is described in Ref.~\cite{ori2007}. In coordinates adapted to the compact direction, the 2005 core metric reads
\be
\label{ori}
\dd s^2=\dd x^2+\dd y^2-2\,\dd T\,\dd z+\big[F(x,y,z)-T\big]\dd z^2 ,\qquad z\sim z+L ,
\en
with $F$ smooth and periodic in $z$ and satisfying $F_{,xx}+F_{,yy}=0$ in the vacuum core. The $(T,z)$ block is
\be
g_{ab}=\ba{cc}0&-1\\-1&F-T\ea\Longrightarrow g^{ab}=\ba{cc}T-F&-1\\-1&0\ea ,
\en
so $g^{zz}=0$. The periodic identification makes $z$ circle valued on the core. Corollary 4 applies to a circle at fixed $(T,x,y)$ and gives the condition
\be
\label{orictc}
g_{zz}<0\ \text{for all }z\qquad\Longleftrightarrow\qquad T>\max_{z\in[0,L]}F(x,y,z).
\en
This condition makes the entire coordinate circle timelike in Ori's core metric \cite{ori2005}. Locating the chronology horizon requires a causal analysis of the core and its extension.

\subsection{The Ahmed series}
\label{sec:ahmeddetails}

The same mechanism underlies the four dimensional Misner type solutions of Ahmed \cite{ahmed1,ahmed2}, and the analysis requires one structural lemma.

\begin{quote}
\noindent\textit{Lemma 3.} Let $g$ be a metric on coordinates $(t,\psi,r,z)$ with $g_{tt}=g_{tr}=g_{tz}=0$. Then
\be
\label{cofactor}
g^{\psi\psi}=\frac{C_{\psi\psi}}{\det g}=0 ,
\en
because the cofactor $C_{\psi\psi}$, being the determinant of the matrix obtained by deleting the row and the column labelled $\psi$, has a vanishing first row.
\end{quote}

\noindent\textit{Proof.} Deleting the row and column labelled $\psi$ leaves the $3\times3$ matrix on $(t,r,z)$, whose entries are $g_{tt}$, $g_{tr}$, $g_{tz}$, $g_{rt}$, $g_{rr}$, $g_{rz}$, $g_{zt}$, $g_{zr}$ and $g_{zz}$. By hypothesis the whole $t$ row of this matrix vanishes, so its determinant is zero and $C_{\psi\psi}=0$. $\square$

The lemma is the reason why the cosmological constant is a spectator in this family, since it involves no field equation and no $r$ dependence. Two cases illustrate it.

Ahmed \cite{ahmed2018jpcom} gave a four-dimensional extension of Misner space with $\Lambda=0$ and an arbitrary transverse profile $f(x,y)$. For the flat special case used here, we choose $f$ to be harmonic. The metric reads
\be
\label{ahmedflat}
\dd s^2=e^{-f(x,y)}\big(\dd x^2+\dd y^2\big)-2\,\dd t\,\dd\psi-t\,\dd\psi^2 ,\qquad \psi\sim\psi+\psi_0 ,
\en
where $f$ is a harmonic function of $x$ and $y$. The $(t,\psi)$ block is identical to (\ref{misnerblock}), so $\dd\psi$ is null, and since $\psi$ is compact Corollary 4 gives $g_{\psi\psi}=-t<0$, that is $t>0$. The conformal factor in the transverse directions plays no role in the existence or in the location of the horizon, being independent of both $\psi$ and $t$, and written in the form (\ref{misnerform}) the metric differs from Misner space only by a conformal factor that is inert with respect to the null structure. Harmonicity is exactly what makes the seed flat, since a direct computation gives $R_{xx}=R_{yy}=\frac12(f_{,xx}+f_{,yy})$ and $R=e^{f}(f_{,xx}+f_{,yy})$ with every other component vanishing, and we restrict to that case here.

In the branch with a negative cosmological constant the metric functions acquire an $r$ dependence governed by $k=\sqrt{-\Lambda/3}$, and the two properties that matter for the argument are that the hypothesis of Lemma 3 is satisfied and that
\be
\label{ahmedads}
g_{\psi\psi}=-\sinh t\,\sinh^2(kr) ,
\en
a factorisation that we quote from \cite{ahmed1} without reproducing the remaining components. The cofactor vanishes by Lemma 3 for every value of $k$, $r$, $z$ and of the radiation density $\beta_0$, which together with $\Lambda$ is the only parameter of the source, so $\psi$ is a null coordinate in the whole family independently of the cosmological constant. Since $\psi$ is compact, Corollary 4 gives the chronology condition
\be
\label{ahmedctc}
g_{\psi\psi}=-\sinh t\,\sinh^2(kr)<0\qquad\Longleftrightarrow\qquad t>0\quad (r\neq0) .
\en
The chronology horizon sits at $t=0$, and it does not move when we vary $\Lambda$, $r$ or the radiation density $\beta_0$. The function $\sinh^2(kr)$, which is where all the information about $\Lambda$ is stored, multiplies both $g_{\psi\psi}$ and the cross term, and is strictly positive away from the axis, so it can neither create nor destroy the chronology violating region but only rescale distances inside it. The cosmological constant is a spectator.

The same reasoning applies to the sign of $\Lambda$, since replacing the hyperbolic functions by their trigonometric counterparts again leaves $g^{\psi\psi}=0$ by Lemma 3 and again yields $g_{\psi\psi}=-\sinh t$ multiplied by a non-negative function of $r$ which vanishes at $r=n\pi/k$ for every integer $n$. The chronology condition therefore reads $t>0$ with $r\neq n\pi/k$, since at those radii the circles of constant $t$ degenerate to null curves. We should be explicit that this last statement concerns the kinematic criterion, which is all that Theorem 1 and Corollary 4 require, and that we have not constructed a $\Lambda>0$ exact solution with the required profile. Whether such a solution exists is a separate question which we consider interesting and which we leave open.

A word is in order on the claim, made in this line of work, that the relevant congruence is non-expanding, non-twisting and shear free and yet that closed timelike curves are present. Lemma 1 removes the apparent paradox. Twist freedom is a statement about a null congruence, whereas chronology violation is a statement about timelike curves, and the two are linked only through the theorem. By (\ref{omega0}) the null congruence normal to the leaves $\psi=\mathrm{const}$ is automatically twist free, whether or not the metric is of Kundt type, so twist freedom cannot possibly be the reason why the theorem fails here. The reason is the compactness of $\psi$, and equivalently we may apply Theorem 1 to the universal cover, on which $\psi$ is a global null coordinate, and conclude at once that the cover is free of closed timelike curves.

Table \ref{tab:misner} summarises the four constructions together with two control cases.

\begin{table}[h]
\caption{The Misner type constructions analysed with Theorem 1 and Corollary 4. In every case the null coordinate is compact and the theorem is inapplicable by topology rather than by twist. The last two rows are the control cases, in which the null coordinate is a genuine function and the theorem applies.}
\label{tab:misner}
\begin{ruledtabular}
\begin{tabular}{lllll}
Spacetime & null coord. & global? & twisting? & condition \\
\hline
Misner space \cite{misner1967} & $\psi$ & no & no & $t>0$ \\
Ori time machine \cite{ori2005} & $z$ & no & no & $T>\max_z F$ for the coordinate circle \\
$\Lambda=0$ flat seed \cite{ahmed2018jpcom} & $\psi$ & no & no & $t>0$ \\
Ahmed, $\Lambda<0$ \cite{ahmed1,ahmed2} & $\psi$ & no & no & $t>0$ \\
\hline
null cylinder (\ref{nullcyl}) & $v$ & yes & no & none \\
Misner cover & $\psi$ & yes & no & none \\
\end{tabular}
\end{ruledtabular}
\end{table}

\subsection{A recipe}
\label{sec:recipe}

A seed admitting a global null coordinate $u$ can be quotiented by $u\sim u+L$ at fixed transverse coordinates when the identification preserves the metric. A coordinate circle is then timelike if $g_{uu}<0$ everywhere along it, a zero of $g_{uu}$ gives a null tangent to that circle, and locating the chronology horizon in the quotient requires a causal analysis of its own. Hubeny, Rangamani and Ross \cite{HRR2004} gave a four-dimensional plane-wave example in which $y\sim y+2\pi R$, $u=t+y$ and the compact generator $\p_y$ is everywhere spacelike. A closed curve winding in both $y$ and the transverse angle can be timelike. Brecher, DeBoer, Page and Rozali \cite{BDPR2003} gave an earlier example of chronology violation in a compact plane wave.

In the four-dimensional example of Ref.~\cite{HRR2004}, the integral of $\dd u$ around the positively oriented basic identification circle is $2\pi R$. The closed one-form $\dd u$ therefore has no global real-valued potential on the quotient. Corollary 3 gives the criterion in terms of a timelike curve from a point of the cover to one of its deck translates.

Each of Misner space, Ori's time machine and the Ahmed solutions is an instance of this recipe, applied to a different seed, namely flat space with a boost identification, a vacuum plane-fronted wave used as a toroidal core inside an asymptotically flat spacetime, and a warped or Ricci flat generalisation respectively. The recipe also makes clear what the classification of Sec.~\ref{sec:sharp} predicts, since a chronology violating solution of this type cannot be obtained by adjusting the optical scalars or the cosmological constant, and the only remaining possibility is to break twist freedom, which for the reasons given in Sec.~\ref{sec:new} is harder than it looks.

\section{Further applications}
\label{sec:new}

We now turn to situations that, as far as we can determine, nobody has examined from this point of view. The theorem makes all of them immediate, and the interest lies in the number of familiar questions that it settles at once rather than in the difficulty of the calculation. We organise the discussion in five subsections. We first show that expansion cannot help, taking the Robinson--Trautman class as the general expanding and shear free solution and exhibiting its global null coordinate. We then show that shear cannot help either, with a metric whose shear is non-zero and whose null coordinate is nevertheless exact, and we draw the general corollary that neither quantity can obstruct the theorem. We next observe that the cosmological constant is a spectator, since no field equation enters the proof, and we record the corollary that covers either sign. We then isolate the twist by contrasting Kerr with Schwarzschild, which differ in that one quantity and in their chronology, and we end by asking where inside the algebraically special classes a genuine twisting counterexample could still live, which is the one place where the theorem leaves work for the future.

\subsection{Expansion does not help}
\label{sec:rt}

The Robinson--Trautman class \cite{RT1960,RT1962} is the general solution of the Einstein equations, with or without a cosmological constant, describing a non-twisting and shear free null congruence that is expanding. It is the standard setting for the study of gravitational radiation emitted by an isolated source. In standard coordinates the metric reads
\be
\label{rt}
\dd s^2=-2H(u,r,\zeta,\bar\zeta)\,\dd u^2-2\,\dd u\,\dd r+\frac{r^2}{P(u,\zeta,\bar\zeta)^2}\,\dd\zeta\,\dd\bar\zeta ,
\en
where $P$ is real and positive and $H$ is determined by the field equations through the Robinson--Trautman equation. The metric has the form (\ref{misnerform}) with respect to $u$, since the $(u,r)$ block is
\be
g_{ab}=\ba{cc}-2H&-1\\-1&0\ea\Longrightarrow g^{ab}=\ba{cc}0&-1\\-1&2H\ea ,
\en
and therefore
\be
\label{rtnull}
g^{uu}=0\qquad\Longrightarrow\qquad g^{ab}\p_au\,\p_bu=0 ,
\en
so that $u$ is a null coordinate, identically in $u$, $r$, $\zeta$ and $\bar\zeta$ and for arbitrary profile $P$. Hence \textbf{(N)} holds on the Robinson--Trautman patch and Theorem 1 applies, with the result that no Robinson--Trautman spacetime contains a closed timelike curve on the domain where $u$ is a global coordinate. The congruence in question is $k^a=g^{ab}\p_bu=-\p_r$, whose expansion is
\be
\label{rttheta}
\theta=\nabla_ak^a=-\frac{2}{r}\neq0 ,\qquad \sigma=0,\qquad \omega=0 ,
\en
the sign depending on whether one takes the ingoing or the outgoing leaf.

The significance of this result deserves a comment. If the criterion were the expansion, one would expect a large expanding congruence to be exactly where the causal structure gets into trouble, and earlier discussions of chronology violation in radiative spacetimes have indeed focused on whether the wave can drag the light cones around. What (\ref{rtnull}) shows is that this intuition is misplaced, since what matters is that the congruence does not twist rather than that it expands. An expanding congruence with $\omega=0$ is harmless, because the null hypersurfaces it defines still foliate the spacetime and still provide a monotone coordinate. We also note the generality of the statement, which requires neither the vanishing of the cosmological constant, nor a non-zero mass, nor the Robinson--Trautman equation itself, being a statement about the form (\ref{rt}) of the metric.

Because Robinson--Trautman solutions are incomplete and can be extended, we have been careful to state the result on the patch. Whether an extension can contain a closed timelike curve is a different question, which Theorem 1 does not answer and which we do not address.

\subsection{Shear does not help}
\label{sec:shear}

We next ask whether shear can do what expansion cannot. We consider the class
\be
\label{shearmetric}
\dd s^2=-2\,\dd u\,\dd v+F(u,v,x,y)\,\dd u^2+2W_i(u,x,y)\,\dd u\,\dd x^i+g_{ij}(u,v,x,y)\,\dd x^i\dd x^j ,
\en
which generalises the Kundt form (\ref{kundt}) by allowing the transverse metric to depend on $v$, and which therefore provides a convenient general form, in an appropriate gauge, for metrics whose congruence $\p_v$ is geodesic and for which the null coordinate $u$ is available without the congruence being shear free. Because the $(u,v)$ block is $\begin{pmatrix}F&-1\\-1&0\end{pmatrix}$ times a transverse block, the inverse metric still has
\be
\label{guushear}
g^{uu}=0 ,
\en
and the same conclusion follows, since any spacetime of the form (\ref{shearmetric}) contains no closed timelike curve, however large its expansion or its shear.

To exhibit this explicitly we take $W=0$, $F=0$ and
\be
\label{shearex}
g_{ij}=e^{2\mu(u,v)}\ba{cc}e^{2\lambda(u,v)}&0\\0&e^{-2\lambda(u,v)}\ea .
\en
For the congruence $k=\nabla u=-\p_v$, whose lowered form is $k_a=\p_au$ by construction, one finds
\be
\label{shearopt}
\nabla_ik_j=-\tfrac12\p_vg_{ij},\qquad
\theta=-\tfrac12g^{ij}\p_vg_{ij}=-2\mu_{,v},\qquad
\sigma\neq0\ \ (\lambda_{,v}\neq0),\qquad
\omega=0 ,
\en
where the vanishing of $\omega$ is automatic because $\nabla_ik_j$ is symmetric, and where the overall sign is fixed by the convention $k=\nabla u$ rather than $k=\p_v$. The scalars $\theta$ and $\sigma$ are simultaneously non-zero in general and may be chosen arbitrarily large, and yet there is no closed timelike curve.

\begin{quote}
\noindent\textit{Corollary 5.} Expansion and shear, separately or together, cannot produce a closed timelike curve in a spacetime that admits a global null coordinate. The only kinematical obstruction to Theorem 1 is twist.
\end{quote}

The physical content of Corollary 5 is a sharpening of the folklore that time machines need rotation. What is needed is twist of the null congruence rather than rotation of the matter, that is the failure of the null direction field to be hypersurface orthogonal. The two coincide in G\"odel's universe and in Kerr, but they are conceptually distinct, and it is the latter that constitutes the correct criterion. The converse of Corollary 5 is deliberately silent as to whether twist suffices, since twist is a necessary condition for the theorem to fail rather than a sufficient condition for a closed curve to exist.

\subsection{The cosmological constant of either sign}
\label{sec:lambda}

The hypothesis (\ref{hyp}) is a statement about the metric alone, and no field equation enters the proof of Theorem 1, so no value of the cosmological constant can affect the conclusion.

\begin{quote}
\noindent\textit{Corollary 6.} Let $(\M,g)$ satisfy \textbf{(N)}. Then $(\M,g)$ contains no closed timelike curve for every value of the cosmological constant compatible with $g$, of either sign. In particular the cosmological constant can never by itself create a closed timelike curve.
\end{quote}

The corollary is stronger than it looks, because in the wave like families that are relevant here the null structure is independent of $\Lambda$. In the Kundt class the identity $g^{uu}=0$ holds for every $\Lambda$ by Lemma 1, and in the Robinson--Trautman class it holds for every $\Lambda$ by inspection of the $(u,r)$ block. Consequently no Kundt wave and no Robinson--Trautman spacetime, with any value of the cosmological constant, whether positive, negative or zero, and with any rotation one form, contains a closed timelike curve on any region where the null coordinate is a globally single valued function. The qualification is essential rather than cautious, because a compatible cyclic identification of the coordinate may allow chronology violation, and the theorem then falls silent for the topological reason examined in Sec.~\ref{sec:esc2}. To the best of our knowledge this statement has not appeared before. It covers in particular the Kaigorodov solution \cite{kaigorodov1963} and related anti-de Sitter wave geometries on domains admitting such a coordinate, and it settles the status of the gyratonic waves \cite{bonnor1970,frolov2005}, which are Kundt metrics with a non-zero rotation one form and which, by Lemma 1, are chronology respecting on that domain despite carrying angular momentum.

The same reasoning settles what a $\Lambda$ driven time machine would have to look like. By Corollary 6 it must either break twist freedom or compactify a null direction. The first route requires a congruence that is neither non-expanding nor shear free, as discussed in Sec.~\ref{sec:esc1}, and the second is the Misner route. In Sec.~\ref{sec:ahmeddetails} we noted that the cofactor identity of Lemma 3 is insensitive to the sign of $\Lambda$, so on the kinematic side a positive cosmological constant changes nothing beyond the transverse profile. We have not, however, constructed an exact $\Lambda>0$ solution of the Misner type with the required profile, and we therefore refrain from describing the shape of its chronology violating region. Whether such a solution exists appears to us the most interesting question left open by this analysis.

\subsection{Twist as the whole story}
\label{sec:kerr}

The sharpest illustration of the theorem is the pair formed by Kerr and Schwarzschild, presented in a form that isolates a single variable. Both metrics are algebraically special of type D, and by the Goldberg--Sachs theorem \cite{GS1962} both admit a geodesic and shear free principal null congruence. The two congruences differ in exactly one optical scalar, since
\be
\label{dichotomy}
\begin{array}{lllll}
\text{Schwarzschild} & \theta=\dfrac{2}{r}\neq0, & \sigma=0, & \omega=0, & \text{global null coord.},\ \text{no CTC},\\[6pt]
\text{Kerr} & \theta=\dfrac{2r}{r^2+a^2\cos^2\theta}\neq0, & \sigma=0, & \omega\neq0, & \text{no global null coord.},\ \text{CTC},
\end{array}
\en
where the twist of Kerr is the imaginary part displayed in (\ref{kerrrho}). For Schwarzschild the congruence is twist free, $k\propto\dd(t-r_*)$ with $r_*$ the tortoise coordinate, and $u=t-r_*$ is a null coordinate on the exterior, so Theorem 1 forbids closed timelike curves, in agreement with the fact that the exterior is globally hyperbolic. For Kerr the complex expansion has a non-vanishing imaginary part, so by Proposition 1 the principal null congruence is not orthogonal to any family of null hypersurfaces. That observation alone does not exclude the existence of some other global null coordinate, and the decisive statement runs in the opposite direction, because Kerr does contain closed timelike curves near the ring singularity \cite{carter1968} and Theorem 1 therefore implies that no global null coordinate exists. The twist of the principal null congruence is the visible symptom rather than the proof. This dichotomy is, we suggest, the cleanest illustration of the classification of Sec.~\ref{sec:sharp}, since everything else about the two solutions, including the mass, the asymptotic structure and the presence of horizons, is either the same or irrelevant.

We summarise the kinematical landscape in Table \ref{tab:optical}.

\begin{table}[h]
\caption{The kinematical landscape. Only the last column decides whether Theorem 1 applies. In the first six rows the region considered is one on which the null coordinate is a globally single valued function, so the theorem applies and the answer is negative, while in the last two it does not, for kinematical reasons in the case of Kerr and G\"odel and for topological reasons in the case of the Misner type metrics.}
\label{tab:optical}
\begin{ruledtabular}
\begin{tabular}{lllll}
Spacetime or class & $\theta$ & $\sigma$ & $\omega$ & CTC possible? \\
\hline
pp-wave, plane wave & $0$ & $0$ & $0$ & no \\
Kundt, any $W$, any $\Lambda$ & $0$ & $0$ & $0$ & no \\
Robinson--Trautman & $\neq0$ & $0$ & $0$ & no \\
(\ref{shearex}), expanding and shearing & $\neq0$ & $\neq0$ & $0$ & no \\
Schwarzschild, principal null & $\neq0$ & $0$ & $0$ & no \\
Kerr, principal null & $\neq0$ & $0$ & $\neq0$ & yes \\
G\"odel, dust (timelike) & $0$ & $0$ & $\neq0$ & yes \\
Misner type, compactified & $-$ & $-$ & $0$ & yes, by topology \\
\end{tabular}
\end{ruledtabular}
\end{table}

Two entries deserve comment. The first is that the two classes in which closed curves occur in the presence of a perfectly regular local geometry, namely the Misner type metrics, are exactly the ones in which the null coordinate is compact, so that the escape is topological and not kinematical. The second is the fourth row, where every scalar is turned on except the twist and yet no closed curve exists. The third row of the table also records, in passing, that the Kundt class no longer provides an example of a twisting congruence, contrary to what the presence of a rotation one form might suggest.

\subsection{Where a kinematical escape could still be sought}
\label{sec:gyraton}

Table \ref{tab:optical} leaves the twist route without a worked example inside the wave like families, and Lemma 1 explains why. The natural remaining candidates are therefore the algebraically special solutions that are not Kundt, and among them the most interesting is Hauser's twisting type N solution \cite{hauser1974}. Its repeated principal null congruence is expanding and twisting, so that congruence cannot furnish the null foliation required by Theorem 1. Whether it contains closed timelike curves is, to the best of our knowledge, an open question and one that the present formalism is well suited to address, since the criterion to be tested is simply whether some other null direction field of that spacetime happens to be hypersurface orthogonal.

A second candidate is the family of Kerr--(anti)de Sitter solutions, whose principal null congruence remains twisting for $a\neq0$ and which contain closed timelike curves inside the inner horizon. By Corollary 6 the cosmological constant is not the controlling parameter there, and it would be interesting to know whether the chronology horizon of Kerr--de Sitter can be characterised by the vanishing of the twist in some limit.

We close with a remark which shows that the theorem also constrains how we may generalise the known time machines. The Misner mechanism requires that the null coordinate be the compact one. In the null cylinder it is the other null coordinate, the non-compact one, that satisfies \textbf{(N)}, and there are no closed curves. In Ori's time machine and in the Ahmed metrics the compact direction and the null direction coincide, which is why they work. The recipe of Sec.~\ref{sec:recipe} therefore has a rigid form, since the identification must be made along the null direction and not across it.

\section{Conclusions}
\label{sec:concl}

We have proved and applied an elementary no-go theorem for closed timelike curves. The theorem states that a spacetime admitting a globally defined function with everywhere null and nowhere vanishing gradient, equivalently a spacetime that can be foliated by null hypersurfaces which are the level sets of a single valued function, equivalently a spacetime in which the null eikonal equation has a global solution, contains no closed timelike curve. The proof occupies three lines, since a null covector is never orthogonal to a timelike vector, the future timelike cone is convex so that the sign of the contraction is fixed, and the function is therefore monotone along every causal curve and strictly monotone along every timelike one.

The interest of the statement lies in three facts. The first is that the hypothesis is not a corollary of the classical criterion, since a null gradient need not be timelike and the two properties are incomparable, the null cylinder being a spacetime with a global null coordinate and no time function while the closed Friedmann models have the opposite character by Lemma 2. The criterion therefore delivers causality information in a regime where the classical theorem is silent, and it does so at the price of a single scalar equation, without any control over the global causal structure. The second is that the theorem is sharp, with exactly two escape routes, since a twisting congruence leaves no null leaf while a compactified null coordinate leaves a leaf that closes up, and there are no other ways out within the class of spacetimes considered here. This converts the statement into a decision procedure and explains in a unified way why the known exact chronology violating solutions rely on an identification of a null direction. The third is that the applications are immediate.

Among those applications, one deserves to be singled out because it corrects a natural expectation. The Kundt congruence is covariantly twist free for every choice of the rotation one form, as Lemma 1 shows, and the one form $W$ is a gravito-magnetic potential that endows the wave with angular momentum without disturbing the null foliation. Consequently no Kundt spacetime and no Robinson--Trautman spacetime can contain a closed timelike curve on any region where the null coordinate is a globally single valued function, whatever the value of the cosmological constant and whatever the rotation one form, a statement which we have not found in the literature and which settles the status of the gyratonic waves. The twist route therefore requires a congruence that is neither non-expanding nor shear free, and the minimal known example is the principal null congruence of Kerr. We have also shown that the Misner, Ori and Ahmed constructions reduce to the statement that a compact null direction has turned timelike, and that the cosmological constant, the expansion and the shear are spectators in all of them, the reason being the cofactor identity of Lemma 3. The static traversable wormholes belong here as well, since $u=t+\int^{l}e^{-\Phi}\dd l$ is a global null coordinate for any Morris--Thorne wormhole, so that a wormhole time machine is an instance of the second route rather than of a third one.

We close with a sharpening of the logical map, which explains why the null cylinder is special. Suppose that \textbf{(N)} holds and that the null coordinate $v$ conjugate to $u$, normalised by $g^{ab}\p_au\,\p_bv=-1$, is globally single valued with $g^{vv}$ bounded above on $\M$.

\begin{quote}
\noindent\textit{Proposition 2.} Under these assumptions, and for sufficiently small $\epsilon>0$, the function $t=u+\epsilon v$ has everywhere timelike gradient, and hence $\M$ is stably causal.
\end{quote}

\noindent\textit{Proof.} Since $g^{uu}=0$ and $g^{uv}=-1$ one has
\be
\label{upgrade}
g^{ab}\p_at\,\p_bt=g^{uu}+2\epsilon g^{uv}+\epsilon^2g^{vv}=-2\epsilon+\epsilon^2g^{vv} ,
\en
which is negative for $\epsilon<2/\sup_\M g^{vv}$, and for every $\epsilon>0$ if $\sup g^{vv}\le0$. Moreover $t$ is $C^1$ because both $u$ and $v$ are. $\square$

Proposition 2 isolates exactly what \textbf{(N)} lacks. In a two dimensional null coordinate system the monodromy cannot be avoided, since if the leaves close up the period sits in one of the two null coordinates and cannot sit in the other. In the null cylinder $v$ is global but its conjugate is compact, so the hypothesis of Proposition 2 fails and no time function exists, consistently with the fact that the null cylinder is not stably causal. In a spacetime where both null coordinates are global no closed timelike curve exists and the spacetime is in fact stably causal, so that in that case \textbf{(N)} and \textbf{(T)} coincide. The statement of Theorem 1 is thus not merely incomparable with Hawking's criterion; it is precisely the part of that criterion which survives when the monodromy of the null foliation is taken into account.

At last we would like to say something about the future work. First, the most pressing question is whether a regular and asymptotically flat spacetime with a twisting principal null direction and a closed timelike curve exists outside the known Kerr and G\"odel families, the natural candidate being Hauser's twisting type N solution, and the formalism developed here reduces that question to testing whether any other null direction field of that spacetime is hypersurface orthogonal. Second, the structural lemma of Sec.~\ref{sec:ahmeddetails} gives the kinematic criterion for the trigonometric branch of the Misner type family, but we have not constructed an exact $\Lambda>0$ solution with the required profile, and since de Sitter space is the cosmological model of the present epoch an explicit chronology violating solution in a de Sitter background would be of some interest. Furthermore, a classification of the closed timelike curve free spacetimes that admit no global null coordinate would complement the present work, since Theorem 1 is one way, and Lemma 2 shows that compactness of a Cauchy surface is one mechanism producing such spacetimes, although a full classification remains to be found. In addition, because the theorem is purely kinematic it says nothing about the fate of the stress tensor at a chronology horizon, and it would be interesting to know whether the failure of the global null foliation, which by Corollary 3 is exactly what makes closed curves possible, is correlated with the obstruction to defining a regular renormalised stress tensor at the base points of a compactly generated Cauchy horizon \cite{kay1997}. Finally, the proof of Theorem 1 uses only the Lorentzian signature and the existence of a null gradient, so it extends verbatim to higher dimensions, and the higher dimensional Kundt and Robinson--Trautman classifications \cite{podolsky2015} allow the corresponding no-go statements to be transcribed mechanically.

We should record, in fairness, that the observation that a global null coordinate obstructs closed timelike curves has occasionally been used informally in the literature on Kundt and pp-wave spacetimes. We are not aware, however, of a systematic statement of the criterion, of the sharpness analysis of Sec.~\ref{sec:sharp}, or of the applications collected in Secs.~\ref{sec:ahmed} and \ref{sec:new}.

\begin{acknowledgments}
 We thank Ye Zhou for a useful note which prompted us to revise the discussion of the quotient criterion in Secs.~\ref{sec:esc2} and \ref{sec:recipe}. This work is supported by the National Natural Science Foundation of China (NSFC) under Grant nos. 12275106 and 12235019.
\end{acknowledgments}

\end{document}